\documentclass[twocolumn]{aastex631}

\DeclareUnicodeCharacter{02BC}{\-}
\shorttitle{NGC4261}
\shortauthors{Yan et al.}

\usepackage{textcomp, gensymb}
\usepackage{booktabs}
\usepackage{soul}
\usepackage{float}

\usepackage{subfigure}
\usepackage{wrapfig}

\begin{document}
\title{Kinematics and Collimation of the Two-Sided Jets in NGC\,4261: VLBI Study on Sub-parsec Scales}
\author[0009-0003-6680-1628]{Xi Yan}
\affiliation{Shanghai Astronomical Observatory, Chinese Academy of Sciences, 80 Nandan Road, Shanghai 200030, People's Republic of China}
\affiliation{School of Astronomy and Space Sciences, University of Chinese Academy of Sciences, 19A Yuquan Road, Beijing 100049, People's Republic of China}

\author[0000-0002-7692-7967]{Ru-Sen Lu}
\affiliation{Shanghai Astronomical Observatory, Chinese Academy of Sciences, 80 Nandan Road, Shanghai 200030, People's Republic of China}
\affiliation{Key Laboratory of Radio Astronomy and Technology, Chinese Academy of Sciences, A20 Datun Road, Chaoyang District, Beijing, 100101, People's Republic of China}
\affiliation{Max-Planck-Institut f\"ur Radioastronomie, Auf dem Hügel 69, D-53121 Bonn, Germany}

\correspondingauthor{Ru-Sen Lu}
\email{rslu@shao.ac.cn}

\author[0000-0001-7369-3539]{Wu Jiang}
\affiliation{Shanghai Astronomical Observatory, Chinese Academy of Sciences, 80 Nandan Road, Shanghai 200030, People's Republic of China}

\author[0000-0002-4892-9586]{Thomas P. Krichbaum}
\affiliation{Max-Planck-Institut f\"ur Radioastronomie, Auf dem Hügel 69, D-53121 Bonn, Germany}

\author[0000-0003-3540-8746]{Zhi-Qiang Shen}
\affiliation{Shanghai Astronomical Observatory, Chinese Academy of Sciences, 80 Nandan Road, Shanghai 200030, People's Republic of China}
\affiliation{Key Laboratory of Radio Astronomy and Technology, Chinese Academy of Sciences, A20 Datun Road, Chaoyang District, Beijing, 100101, People's Republic of China}

\begin{abstract}
We report multi-frequency VLBI studies of the sub-parsec scale structure of the two-sided jet in the nearby radio galaxy NGC\,4261. Our analyses include new observations using the Source Frequency Phase Referencing technique with the Very Long Baseline Array at 44 and 88\,GHz, as well as archival data at 15 and 43\,GHz. Our results show an extended double-sided structure at 43/44\,GHz and provide a clear image of the nuclear region at 88\,GHz, showing a core size of $\sim$0.09 mas and a brightness temperature of $\sim1.3\times10^{9}$\,K. Proper motions are measured for the first time in the two-sided jet, with apparent speeds ranging from $0.31\pm0.14\,c$ to $0.59\pm0.40\,c$ in the approaching jet and $0.32\pm0.14\,c$ in the receding jet. The jet-to-counter-jet brightness ratio allows us to constrain the viewing angle to between $\sim54\degree$ and $84\degree$ and the intrinsic speed to between $\sim0.30\,c$ and $0.55\,c$. We confirm the parabolic shape of the upstream jet on both sides of the central engine, with a power-law index of $0.56\pm0.07$. Notably, the jet collimation is found to be already completed at sub-parsec scales, with a transition location of about 0.61 pc, which is significantly smaller than the Bondi radius of 99.2 pc. This behavior can be interpreted as the initial confinement of the jet by external pressure from either the geometrically thick, optically thin advection-dominated accretion flows (ADAF) or the disk wind launched from it. Alternatively, the shape transition may also be explained by the internal flow transition from a magnetically dominated to a particle-dominated regime.
\end{abstract}
\keywords{galaxies: active --- galaxies: individual (NGC\,4261) --- galaxies: nuclei --- radio continuum:
galaxies}

\section{Introduction}\label{sec:Introduction}

\begin{deluxetable*}{ccccccccc}
\tablecaption{Summary of NGC\,4261 observations\label{tab:observation summary}}
\tablehead{\colhead{Freq.} & \colhead{P.C.} & \colhead{Date} & \colhead{Array} & \colhead{Pol} & \colhead{Bandwidth} & \colhead{Beam size} & \colhead{ $I_{\rm peak}$  } & \colhead{$I_{\rm rms}$} \\
(GHz) &&&&&(MHz)& (mas $\times$ mas, deg) &(Jy beam$^{-1}$) &(Jy beam$^{-1}$)}
\startdata
15 &BM166 & 2002.07.05 & VLBA  & Dual& 64 & 0.922$\times$0.512, -4.3 & 0.129 & 0.0005 \\
15 &BM175b & 2002.09.27 & VLBA & LCP & 64 & 1.04$\times$0.498, -5.1 & 0.133 & 0.0005\\
15 &BM175c & 2003.05.05 & VLBA & LCP & 64 & 1.01$\times$0.456, -5.12 & 0.121 & 0.0005 \\
15 &BM175a & 2003.07.04 & VLBA & LCP & 64 & 1.02$\times$0.459, -4.66 & 0.130 & 0.0005  \\
\hline
43 & BM215a & 2004.12.20 &VLBA & Dual & 64 & 0.344$\times$0.175, -8.31 & 0.143&0.0006\\
\hline
44 & BY167 & 2022.02.14 & VLBA, -SC,-HN & Dual & 1024 &  0.627$\times$0.171, -22.4 & 0.113 & 0.0005 \\
88 & BY167 & 2022.02.14 & VLBA, -SC,-HN & Dual & 1024 & 0.467$\times$0.101, -19.2 & 0.0492 & 0.0015 \\
\enddata
\tablecomments{
Column (1): Observing frequency.
Column (2): Project code.
Column (3): Date of observation.
Column (4): Participating stations. Stations not involved are indicated with a minus sign.
Column (5): Polarization.
Column (6): Bandwidth.
Column (7): Full width at half maximum (FWHM) and position angle of the synthesized beam.
Column (8)-(9): Peak intensity and rms noise.
}
\end{deluxetable*}

\vspace{-1cm}
Relativistic jets in active galactic nuclei (AGN) undergo poorly understood acceleration and collimation processes that are closely linked to their launching mechanisms. Theoretical studies and simulations \cite[e.g.,][]{McKinney_2006MNRAS.368.1561M,Tchekhovskoy_2011MNRAS.418L..79T} suggest that jets can originate from either a spinning black hole \citep{blandford_1977MNRAS.179..433B} or an accretion flow \citep{blandford_1982MNRAS.199..883B}. Moreover, the initial jet is suggested to be magnetically dominated with a parabolic shape due to external pressure \citep[e.g.,][]{McKinney_2012MNRAS.423.3083M}. However, as the jet propagates, it transits to a kinetically dominated state, expanding freely in a conical shape.

Very Long Baseline Interferometry (VLBI) is a powerful tool for studying the jet formation, acceleration and collimation processes. It has been extensively applied to several nearby low-luminosity AGN (LLAGN) to study jet collimation, such as M\,87 \citep[e.g.,][]{Asada_2012ApJ...745L..28A,Lu_2023}, NGC\,6251 \citep{Tseng_2016ApJ...833..288T}, NGC\,4261 \citep{nakahara_2018ApJ...854..148N}, NGC\,1052 \citep{Nakahara_NGC1052_2020AJ....159...14N} and NGC\,315 \citep{park_2021ApJ...909...76P,boccardi_2021A&A...647A..67B}. Recently, \cite{Kovalev_2020MNRAS.495.3576K} proposed that the transition from a parabolic to conical shape may be a common effect in nearby AGN jets based on their analysis of a sample of 367 AGN. They also noted that the transition location does not necessarily coincide with the Bondi radius. NGC\,315 serves as a typical example, where the jet collimation is completed early on sub-parsec scales \citep{boccardi_2021A&A...647A..67B}. This behavior is interpreted as the initial confinement of the jet by the external pressure exerted by either the ADAF or the disk wind launched from it.

Among the above-mentioned sources, the Fanaroff-Riley Class I (FR-I) source, NGC\,4261, deserves particular attention. First, the jet is observed at a large viewing angle of $63\degree\pm3\degree$ \citep{piner_2001AJ....122.2954P} and is double-sided \citep[e.g.,][]{Jones_1997ApJ...484..186J}. Second, precise core-shift measurements have determined the location of the central supermassive black hole \citep[SMBH, at a distance of $82\pm16\,\mu$as from the 43\,GHz core,][]{haga2015ApJ...807...15H}. This allows an accurate estimate of the de-projected radial distance between the jet and the central SMBH. Furthermore, the proximity of NGC\,4261 \citep[31.6 Mpc,][]{Tonry_2001ApJ...546..681T} and its large black hole mass \citep[$1.62\times10^{9}M_{\sun}$,][]{Boizelle_2021ApJ...908...19B,ruffa_2023MNRAS.522.6170R} make it a valuable laboratory for studying jet properties, with 1 mas corresponding to 0.15 pc or 988 Schwarzschild radii ($R_{\rm s}$).

Despite these advantages, the collimation and kinematics of the NGC\,4261 jet remain largely unexplored. Although previous observations found parabolic-to-conical transition signatures on the jet width profile, the upstream parabolic shape could not be well sampled due to the limited number of width measurements \citep[see Figures 2-4 in][]{nakahara_2018ApJ...854..148N}. In addition, apart from the work by \cite{piner_2001AJ....122.2954P}, who provided only one jet speed measurement, there have been no further kinematic analyses conducted on the NGC\,4261 jet. For these reasons, we aim to examine the width profile of the upstream jet and investigate its kinematics.

This paper is organized as follows. In Section\,\ref{sec:Observations and Data Reduction}, we present our observations and data reduction. Section\,\ref{sec:Data analysis} describes the methods used for our kinematic analysis and transverse width measurement. The results are presented in Section\,\ref{sec:Results}, followed by a discussion in Section\,\ref{sec:Discussions}. Finally, we summarize in Section\,\ref{sec:summary}.

\section{Observations and Data Reduction} \label{sec:Observations and Data Reduction}

\subsection{New VLBA observations}
We observed NGC\,4261 using the Very Long Baseline Array (VLBA) with the Source Frequency Phase Referencing (SFPR) technique \citep{Rioja_2011AJ....141..114R} on February 14, 2022. The observations were performed at 44 and 88\,GHz, with a data rate of 4\,Gbits/s and 2-bit sampling. Both left-hand circular polarization (LCP) and right-hand circular polarization (RCP) were recorded, covering a total bandwidth of 1024\,MHz. Each polarization was divided into 4 sub-bands (IFs). We used 3C\,279 and 3C\,273 as the fringe finder and amplitude/phase calibrator, respectively. A summary of the observations is provided in Table\,\ref{tab:observation summary}.

We calibrated the data using NRAO's Astronomical Image Processing System \citep[AIPS,][]{Greisen_aips_2003ASSL..285..109G} following the procedures in \cite{Jiang_2021ApJ...922L..16J}. The phase calibration involved several steps. Firstly, we removed the constant single-band delays and phase offsets using high signal-to-noise ratio (SNR) calibrator scans. Then, we performed global fringe fitting to eliminate single- and multi-band residual delays and solve for fringe rates. Afterward, we applied frequency phase transfer (FPT) to the 88\,GHz data by multiplying the 44\,GHz residual phase solutions by the frequency ratio of 2. For the 88\,GHz data, a re-fringe-fitting was run on the calibrator 3C\,273 and the solutions were applied to NGC\,4261 to further correct the residual ionospheric errors as well as the instrumental offsets between 44 and 88\,GHz. We performed a prior amplitude calibration using the antenna system temperatures and gain curves with opacity corrections. The band-pass calibration was derived from scans on a bright calibrator source. Once calibration was completed, we averaged the data over frequency and conducted imaging and self-calibration using DIFMAP \citep{Shepherd_difmap_1997ASPC..125...77S}.

\subsection{Archival VLBA data}
We also analyzed archival VLBA data of NGC\,4261 at 15 and 43\,GHz. The details of these observations are provided in Table\,\ref{tab:observation summary}. The BM166 data (15\,GHz) were originally observed for polarimetry \citep{middelberg2004gas}. The three-epoch BM175 datasets were observed at multiple frequencies but we only utilized the 15\,GHz data for our analysis. In addition, we noted that the BM175c data were already published \citep{Middelberg2005A&A...433..897M}. The BM215a data (43\,GHz) were also designed for polarimetry. For all these archival observations, we performed data reduction and imaging using AIPS and DIFMAP following standard procedures \citep[e.g.,][]{Lu_2023}.

\begin{deluxetable}{cccccc}
\tabletypesize{\small}
\tablecaption{Properties of the model-fitted Gaussian components \label{tab:components}}
\tablehead{\colhead{ID} & \colhead{Ep.} & \colhead{$r$ (mas)} & \colhead{$S_{\rm v}$ (mJy)} & \colhead{$d$ (mas)} & \colhead{$\beta_{\rm app}$ (c)}}
\startdata
Core& 1 & 0  & $131\pm13$ &  $0.29\pm0.029$ &    \\
    & 2 & 0  & $126\pm12$ &  $0.27\pm0.026$ &    \\
    & 3 & 0  & $130\pm11$ &  $0.26\pm0.023$ &    \\  
    & 4 & 0  & $133\pm12$ &  $0.26\pm0.022$ &  0 \\
\hline
W1 & 1 & $4.42\pm0.32$ &  $16\pm11$ &    $1.28\pm0.63$ &      \\
   & 2 & $4.75\pm0.26$ &  $10\pm7$  &    $1.13\pm0.52$ &      \\    
   & 3 & $5.33\pm0.50$ &  $15\pm13$ &    $1.89\pm1.00$ &      \\
   & 4 & $5.69\pm0.54$ &  $15\pm14$ &    $2.04\pm1.08$ & $0.59\pm0.40$\\
\hline
W2&1& $2.61\pm0.22$ &  $23\pm11$ &  $1.22\pm0.43$ &    \\
& 2 & $2.75\pm0.22$ &   $26\pm12$ &   $1.20\pm0.44$ &    \\
& 3 & $3.35\pm0.26$ &   $16\pm8$ &    $1.27\pm0.52$ &    \\
& 4 & $3.52\pm0.28$ &   $17\pm9$ &    $1.32\pm0.56$ &   $0.46\pm0.25$ \\
\hline
W3&1& $1.00\pm0.10$ &   $45\pm9$ &    $0.56\pm0.10$ &    \\
& 2 & $1.27\pm0.10$ &   $34\pm8$ &    $0.62\pm0.13$ &    \\
& 3 & $1.97\pm0.11$ &   $19\pm6$ &    $0.76\pm0.22$ &    \\
& 4 & $2.13\pm0.13$ &   $16\pm6$ &    $0.76\pm0.26$ &   $0.57\pm0.12$   \\
\hline
W4&1& $0.46\pm0.10$ &   $88\pm11$ &    $0.31\pm0.10$ &    \\
& 2 & $0.76\pm0.10$ &   $55\pm8$ &    $0.34\pm0.10$ &    \\
& 3 & $1.19\pm0.09$ &   $30\pm6$ &    $0.49\pm0.09$ &    \\
& 4 & $1.42\pm0.09$ &   $21\pm5$ &    $0.50\pm0.11$ &   $0.45\pm0.09$   \\
\hline
W5&2& $0.39\pm0.10$ &   $65\pm8$ &    $0.20\pm0.10$ &     \\
& 3 & $0.76\pm0.09$ &   $55\pm7$ &    $0.31\pm0.09$ &     \\
& 4 & $0.86\pm0.09$ &   $54\pm7$ &    $0.41\pm0.09$ &   $0.31\pm0.14$   \\
\hline
E1&1& $-1.22\pm0.13$ &   $30\pm10$&    $0.87\pm0.26$ &    \\
& 2 & $-1.44\pm0.15$ &   $20\pm8$ &    $0.89\pm0.30$ &    \\
& 3 & $-1.69\pm0.20$ &   $20\pm8$ &    $1.07\pm0.40$ &    \\
& 4 & $-1.96\pm0.20$ &   $12\pm6$ &    $1.00\pm0.40$ &   $0.32\pm0.14$  \\
\enddata
\tabletypesize{\footnotesize}
\tablecomments{
Column (1): Component label. 
Column (2): Epoch\,(1: 2002.07.05, 2: 2002.09.27, 3: 2003.05.05, 4: 2003.07.04). 
Column (3): The radial distance from the core component. Column (4): The flux density. 
Column (5): The size (FWHM). 
Column (6): Apparent speed in units of the speed of light $c$.}
\end{deluxetable}

\section{Data analysis} 
\label{sec:Data analysis}
\subsection{Model fitting} 
\label{sec:model fitting}
\begin{figure}[htbp!]
\begin{center}
\includegraphics[width=0.45\textwidth]{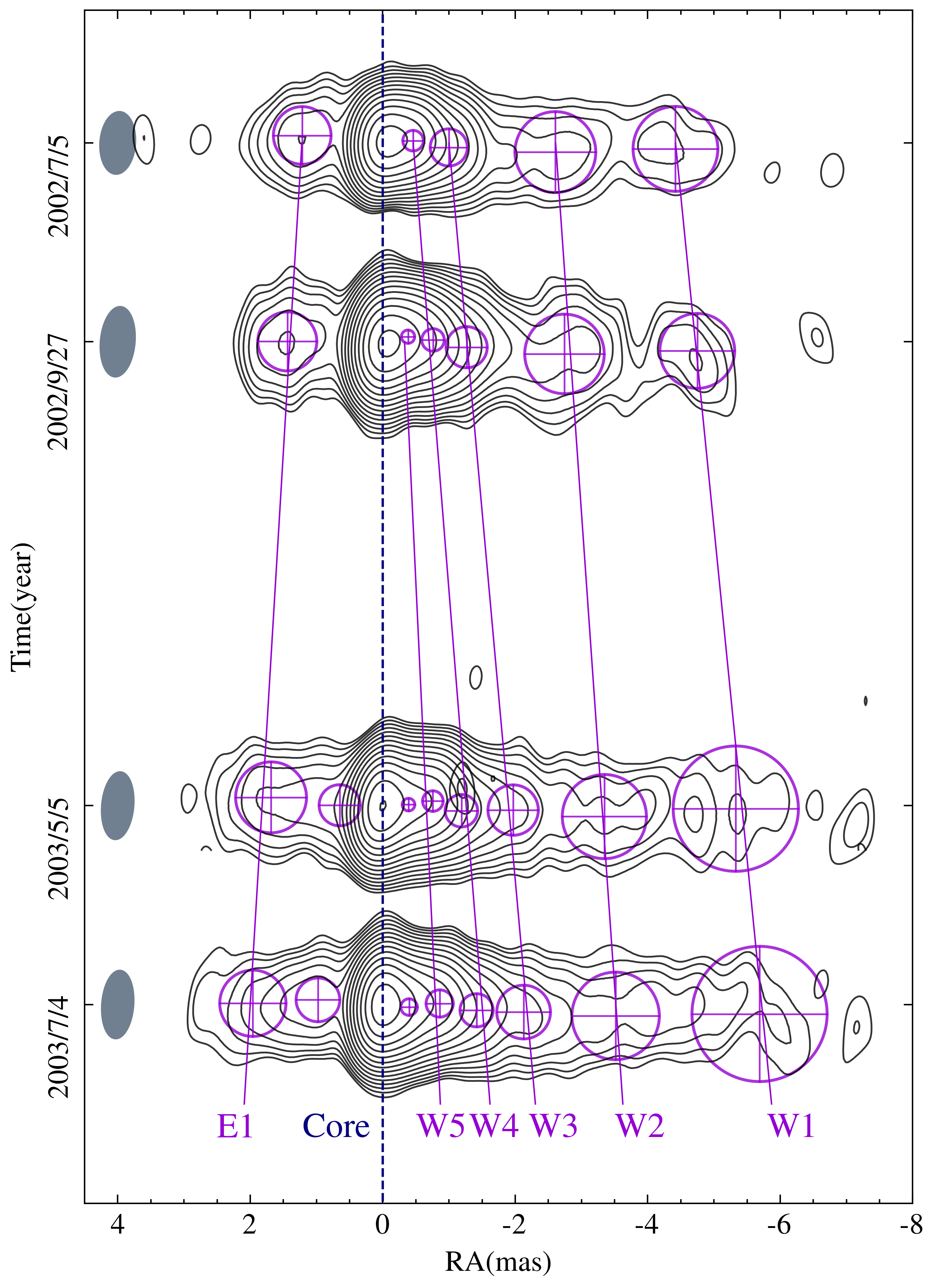}
\caption{Images of NGC\,4261 at 15\,GHz. These images are centered on the bright core position. The fitted circular Gaussian components are represented by dark violet circles superimposed on the contours. The cross-identified components are labeled at the bottom. The dark violet lines depict the best-fit line of proper motion. The slategrey filled ellipses on the left indicate the synthesized beam for each image. Contours begin at 3 times the rms value and increase by a factor of $\sqrt{2}$.}
\label{fig:15G_maps}
\end{center}
\end{figure}

We performed kinematic analysis using the 15\,GHz data. To model the source structure, we fitted several circular Gaussian models to the complex visibilities using the MODELFIT task in DIFMAP. Then the fitted components in the four epochs were cross-identified based on their location, flux density and size (Table\,\ref{tab:components}). To align the images, we used the compact bright core component as the reference position (Figure\,\ref{fig:15G_maps}). The error in the fitted parameters was determined by considering the local SNR in the image around each feature \citep{Lee_2008AJ....136..159L}. For positional uncertainties smaller than one-fifth of the minor beam size, we adopted the latter as the error estimate.

\subsection{Image analysis} \label{sec:transverse structure}
To obtain the transverse structure of the jet, we measured the width of the double-sided jets. For the 15\,GHz data, we used the stacked image created after convolving each individual image with a common beam. As for the 43 and 44\,GHz data, we used the two individual images. Since the jet is almost along the east-west direction, we sliced the jet along PA=$0\degree$ using the AIPS task SLICE and obtained a series of pixel-based transverse intensity profiles. Each transverse intensity profile was fitted with a Gaussian function to determine the full width at half maximum (FWHM), $W_{\rm fit}$. Then we calculated the de-convolved jet width as $W^{2} = W_{\rm fit}^{2} - W_{\rm res}^{2}$, where $W_{\rm res}$ is the resolution along PA=0$\degree$. To obtain the radial profile of the jet width, we calculated the distance from the central engine to each slice location, taking into account the measured core-shift relation \citep{haga2015ApJ...807...15H}.

\begin{figure}[htbp!]
\begin{center}
\includegraphics[width=0.9\columnwidth]{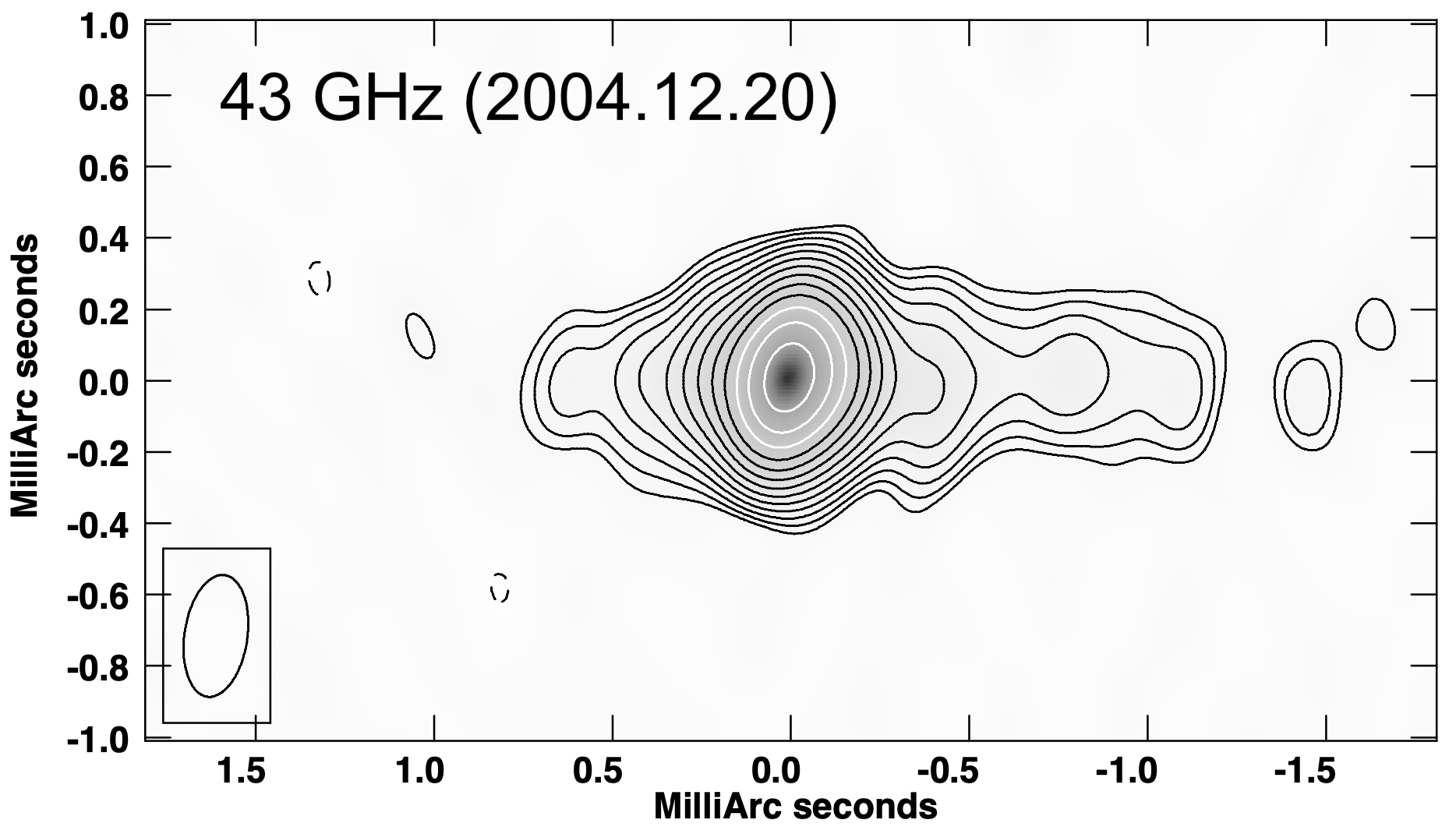}
\includegraphics[width=0.9\columnwidth]{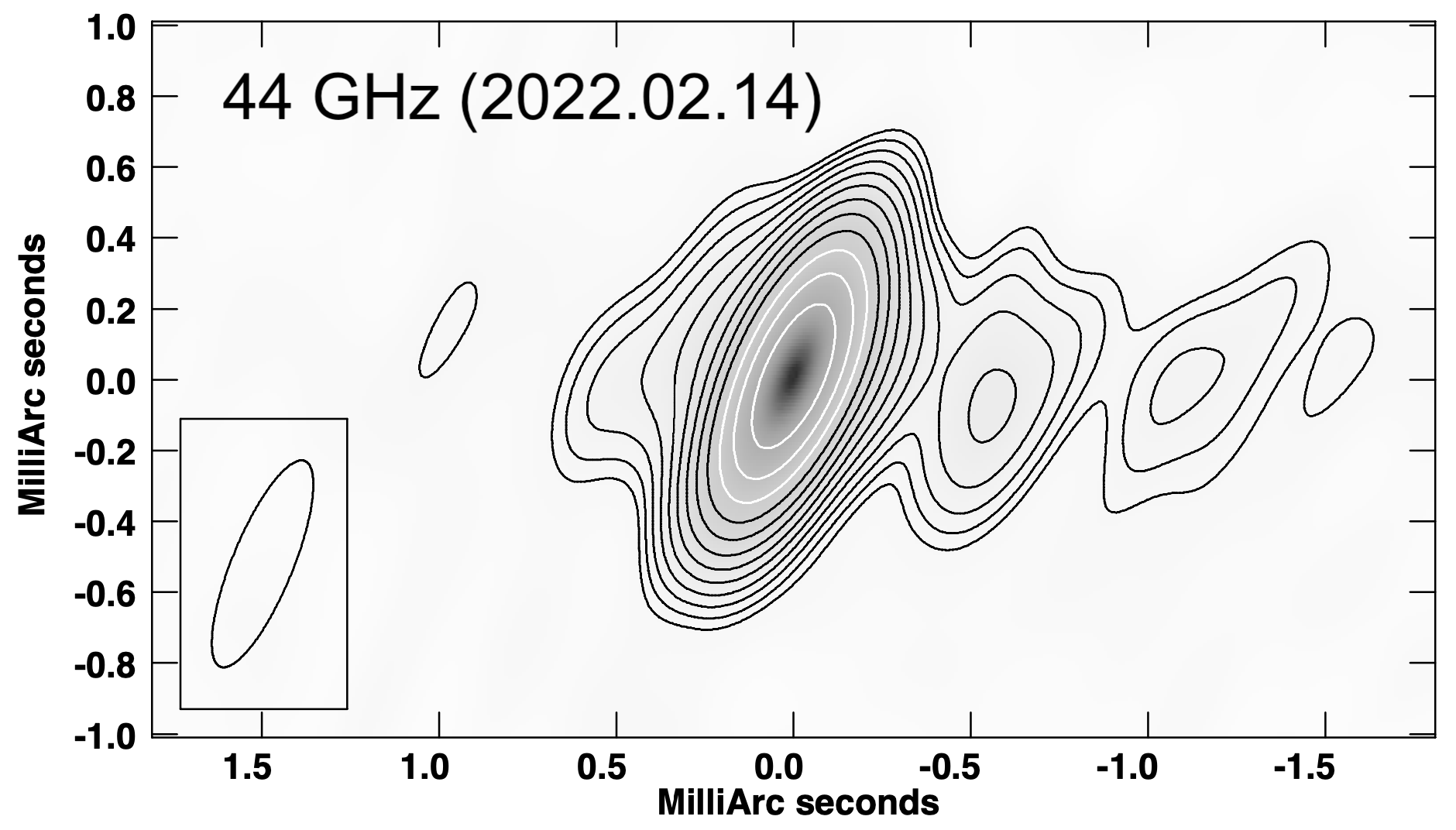}
\includegraphics[width=0.91\columnwidth]{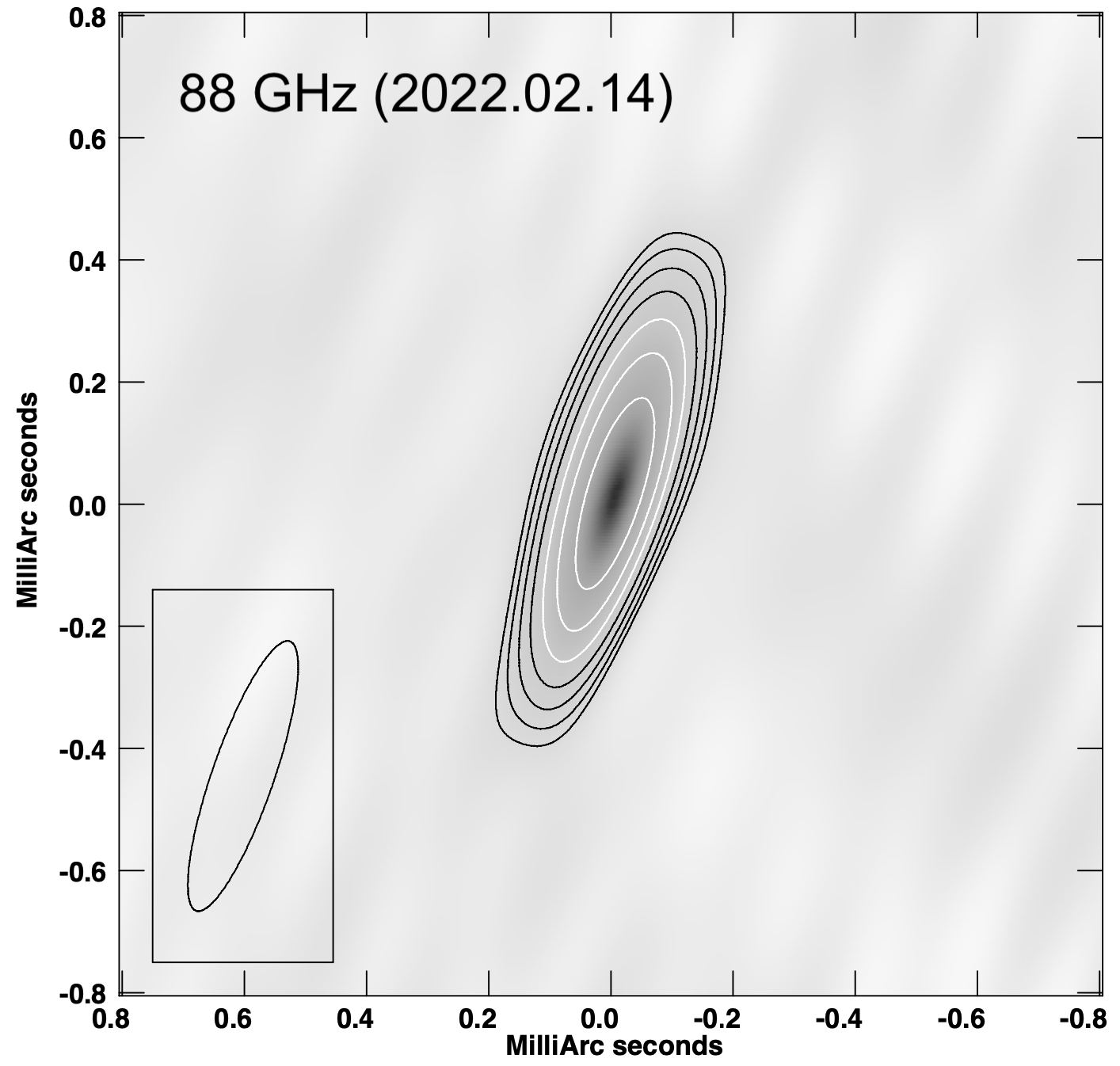}
\caption{Self-calibrated images of the NGC\,4261 jet obtained from VLBA observations at 43, 44 and 88\,GHz. The synthesized beam is shown at the bottom left corner of each image. Contours start at 3 times the rms value and increase by a factor of $\sqrt{2}$.}
\label{fig:43G_maps}
\end{center}
\end{figure}

\begin{figure}[htbp!]
\begin{center}
\includegraphics[width=0.45\textwidth]{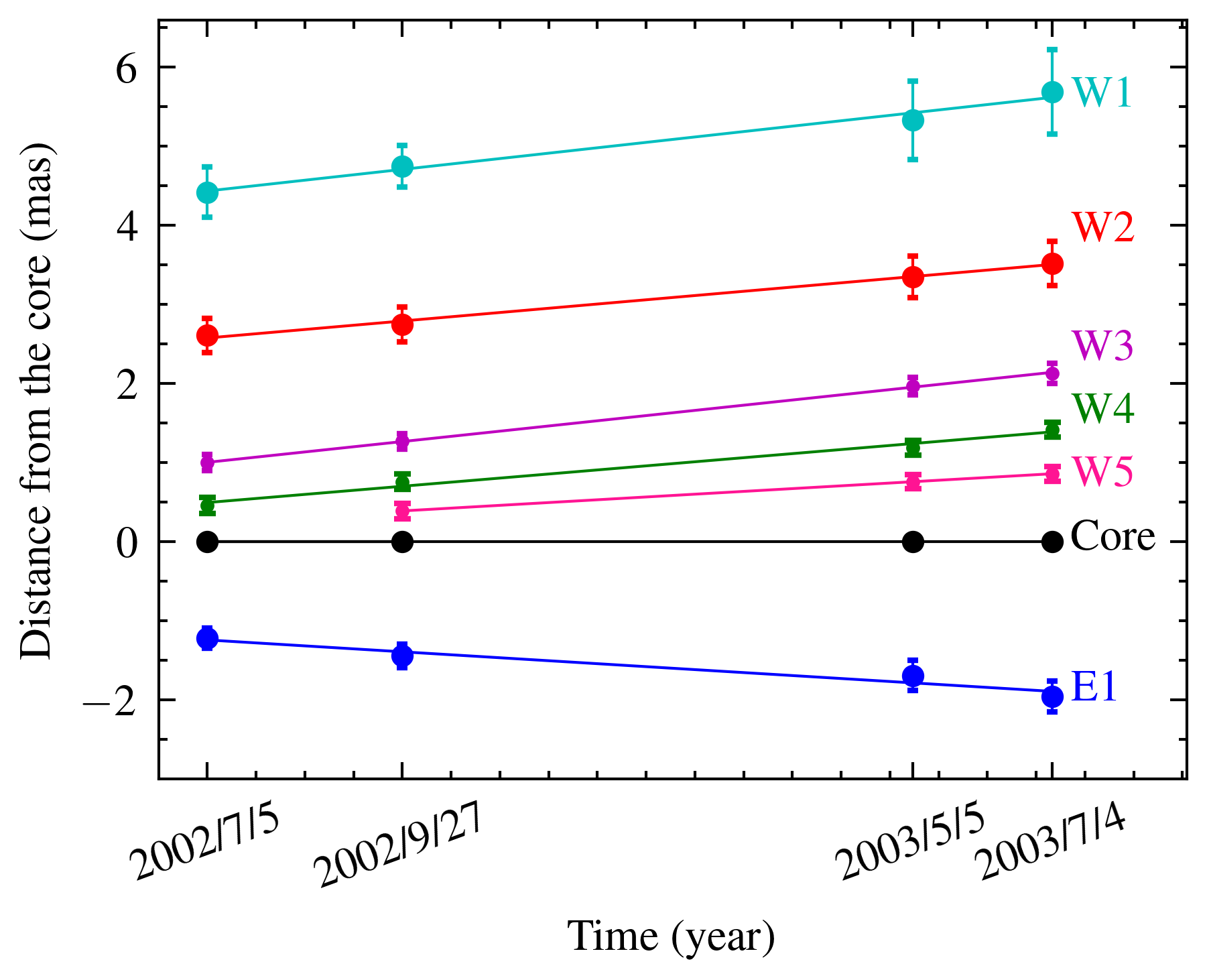}
\caption{Radial distance from the core versus time for the cross-identified components.}
\label{fig:15G_kinematics}
\end{center}
\end{figure}

\begin{figure}[htbp!]
\begin{center}
\includegraphics[width=0.51\textwidth]{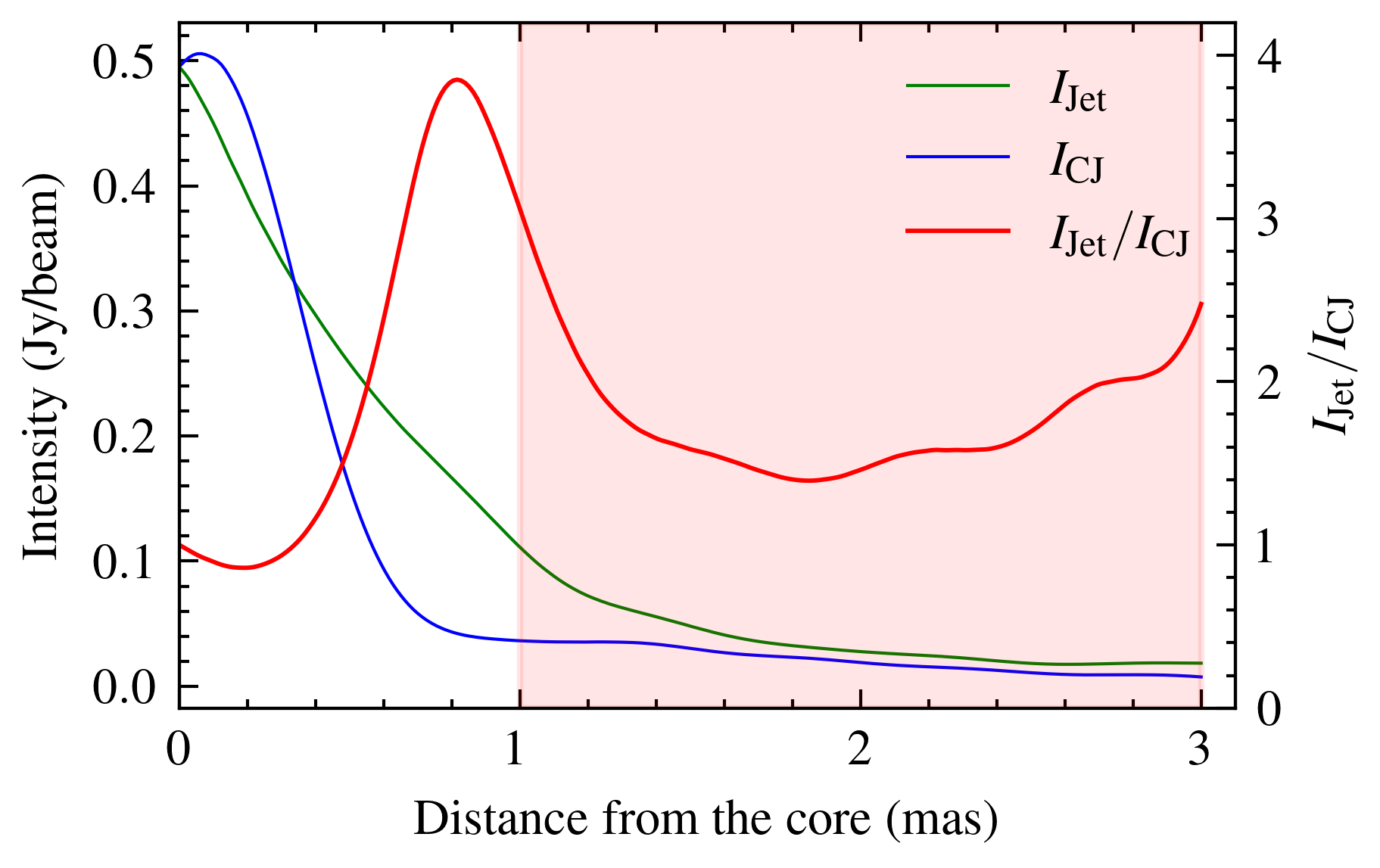}
\includegraphics[width=0.475\textwidth]{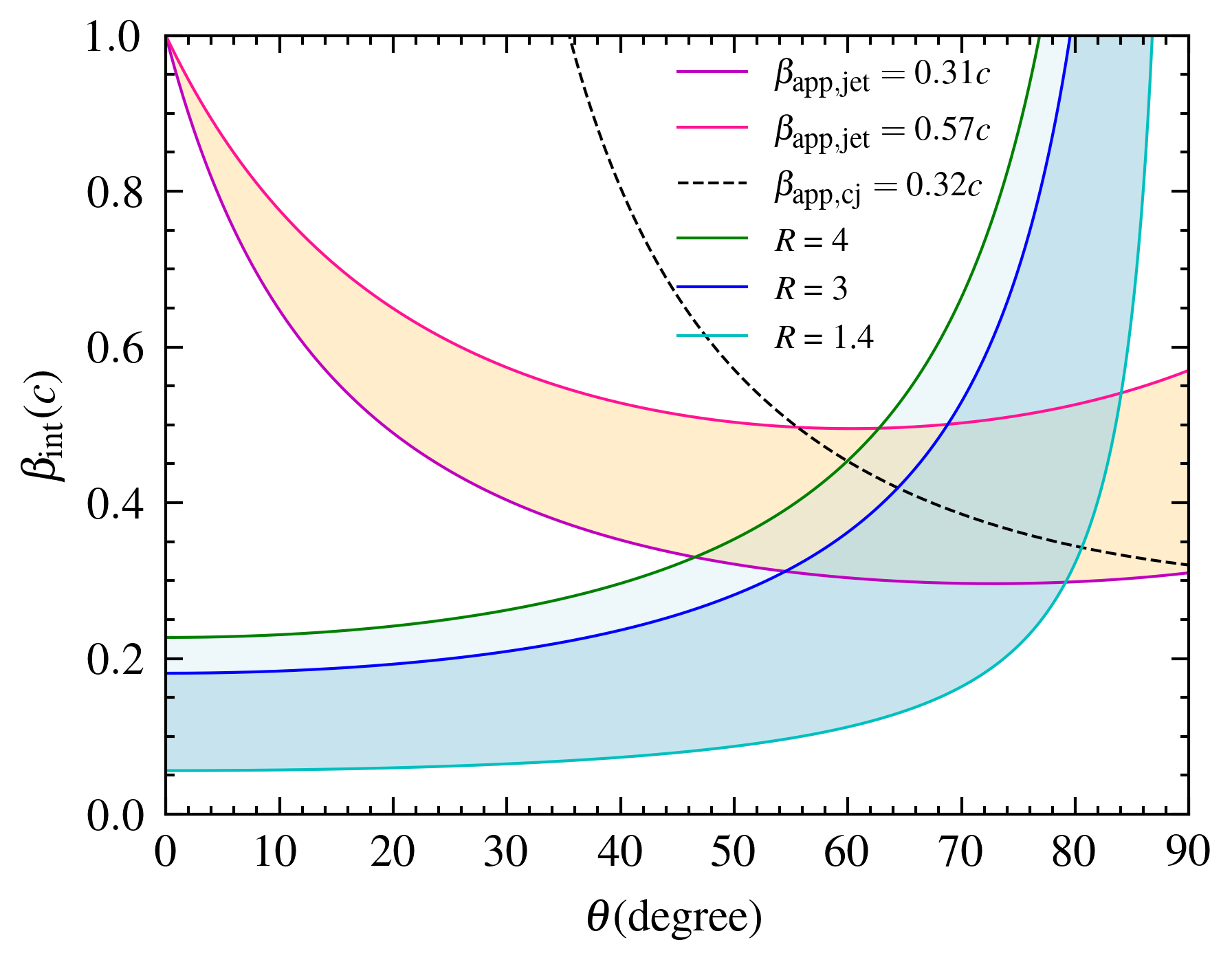}
\caption{Top: The radial intensity profiles of the jet (green) and counter-jet (blue) are shown, along with their corresponding brightness ratio $R$ (red). The brightness ratio within the shaded area was used to constrain the jet viewing angle. Bottom: The allowed range of the viewing angle and intrinsic velocity of NGC\,4261 jet.}
\label{fig:viewing_angle}
\end{center}
\end{figure}

\section{Results} 
\label{sec:Results}
\subsection{Source morphology}
Figures\,\ref{fig:15G_maps} and \ref{fig:43G_maps} show the uniformly weighted CLEAN images of NGC\,4261 jet observed at 15, 43, 44 and 88\,GHz. Clearly two-sided jets were detected at 15, 43 and 44\,GHz, with the western side representing the approaching jet and the eastern side representing the receding jet. 

At 43 and 44\,GHz, we observe a more extended structure compared to previous studies \citep{Jones_2000ApJ...534..165J, Middelberg2005A&A...433..897M}, although the apparent structure are slightly different due to the different beam shape. At 88\,GHz, with an angular resolution of 0.467$\times$0.101 mas, we obtain a clear image of the nuclear structure on a scale as small as 100 $R_{\rm s}$. The derived size of the core is 0.09 mas, from which we estimate a brightness temperature ($T_{\rm B}$) of $1.3\times10^{9}$\,K.

\subsection{Jet kinematics}
Figure\,\ref{fig:15G_maps} displays the measured proper motions of the NGC\,4261 jet. We note that a new component (W5) was ejected between September 2002 and July 2003. By conducting linear fits to the radial distances from the core over time, we determined the apparent speeds of these features (see Figure\,\ref{fig:15G_kinematics} and Table\,\ref{tab:components}). The measured apparent speeds in the approaching jet range from $0.31\pm0.14\,c$ to $0.59\pm0.40\,c$ and in the counter-jet is $0.32\pm0.14\,c$.

The intrinsic velocity ($\beta_{\rm int}$) and the viewing angle ($\theta$) of the jet can be constrained using the apparent velocity ($\beta_{\rm app}$) and the jet-to-counter-jet brightness ratio ($R$). These relationships can be expressed by the following equations:
\begin{equation} \label{eq:beta_int}
\beta_{\rm int} = \frac{\beta_{\rm app}}{\rm sin\theta+\beta_{\rm app}\rm cos\theta}
\end{equation}
and
\begin{equation} \label{eq:brightness_ratio_2}
\beta_{\rm int} = \frac{1}{\rm cos\theta}\left(\frac{R^{1/(2-\alpha)}-1}{R^{1/(2-\alpha)}+1}\right)
\end{equation}
where $\beta_{\rm int}$ and $\beta_{\rm app}$ are in units of $c$, and $\alpha$ represents the spectral index of the jet ($S\propto\nu^{+\alpha}$). We adopted $\alpha = -1$ based on the spectral index map from \cite{haga2015ApJ...807...15H}.

We determined the longitudinal intensity profile along the jet within 3 mas from the core in the stacked 15\,GHz image for both the approaching and receding jet. As shown in the top panel of Figure\,\ref{fig:viewing_angle}, the brightness ratio varies from $\sim$1 to 4. In the same region, we measured the apparent speeds of the approaching jet, which range from 0.31\,$c$ to 0.57\,$c$. By combining these values with the brightness ratios, we were able to constrain the viewing angle to be $\theta\ga46\degree$ (Figure\,\ref{fig:viewing_angle}, bottom).

To measure the brightness ratio of the approaching jet to the receding jet, we excluded the core region to avoid possible biases. This is because the observed central bright core may suffer from blending effects between the base of the approaching and the receding jet, and the emission from the receding jet may also be absorbed by the accretion flow. In doing so, we employed two approaches. First, we excluded the innermost 1 mas region of the flow, which corresponds to twice the minor axis size of the restoring beam \citep[see, e.g.,][]{Mertens_2016A&A...595A..54M}. With this exclusion, the brightness ratio is between $\sim$1.4 and 3 (Figure\,\ref{fig:viewing_angle}, top). This range provides an estimate for the viewing angle of about $54\degree$ to $84\degree$ (Figure\,\ref{fig:viewing_angle}, bottom).

Alternatively, we calculated the brightness ratio by considering the clean components in each individual 15\,GHz map. By placing two rectangular boxes of the same size and distance from the core on both sides of the jet, we obtained a brightness ratio ranging from 1.6 to 2. Additionally, both the 43 and 44\,GHz maps also provided a brightness ratio of about $2$. Overall, these results are all within the range of $1.4\la R\la3$ and point toward a very similar viewing angle range. 

Notably, we also measured an apparent speed of 0.32\,$c$ for the counter-jet at separations from 1 mas to 3 mas. As shown in the bottom panel of Figure\,\ref{fig:viewing_angle}, this apparent speed intersects with the lines given by the measured brightness ratio. These intersections provide a viewing angle range as well: from $\sim64\degree$ (for $R=3$) to $80\degree$ (for $R=1.4$). This is highly consistent with the above analysis using the apparent speeds of the approaching jet. Considering all the above results, we obtain a conservative range of viewing angles from $54\degree$ to $84\degree$ and an intrinsic speed range from $\sim$ 0.30\,$c$ to 0.55\,$c$.

\subsection{The inner collimation profile}
We analyzed the radial width profile of the upstream jet, including measurements at 15, 43 and 44\,GHz (Section\,\ref{sec:transverse structure}). We also considered the 88\,GHz core size as an upper limit for the jet width and estimated its distance to the SMBH to be $\sim$0.036 mas based on the core-shift relation \citep{haga2015ApJ...807...15H}. All measurements were converted to the de-projected physical scales in units of $R_{\rm s}$.

\begin{figure}
\begin{center}
\includegraphics[width=0.48\textwidth]{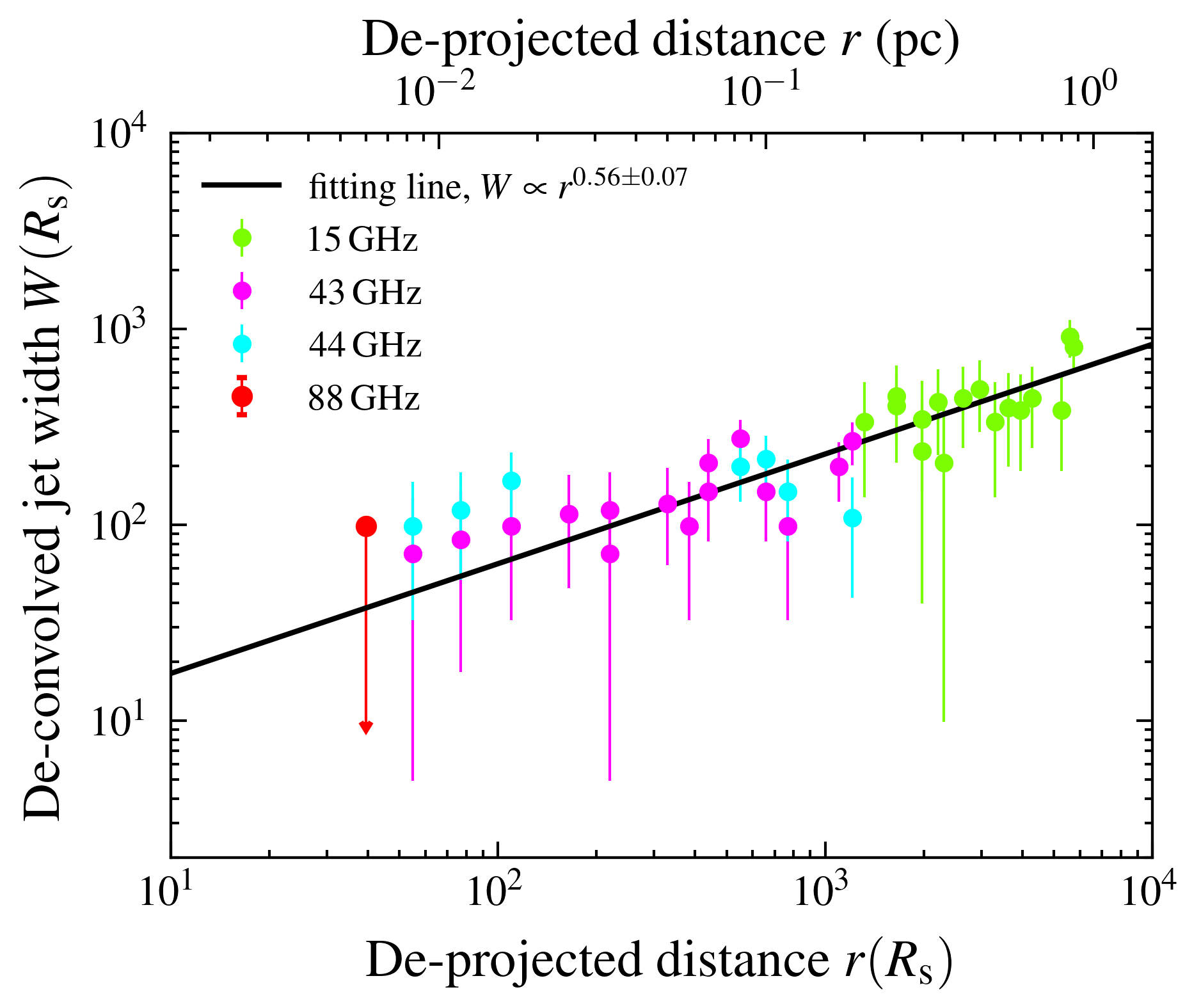}
\includegraphics[width=0.46\textwidth]{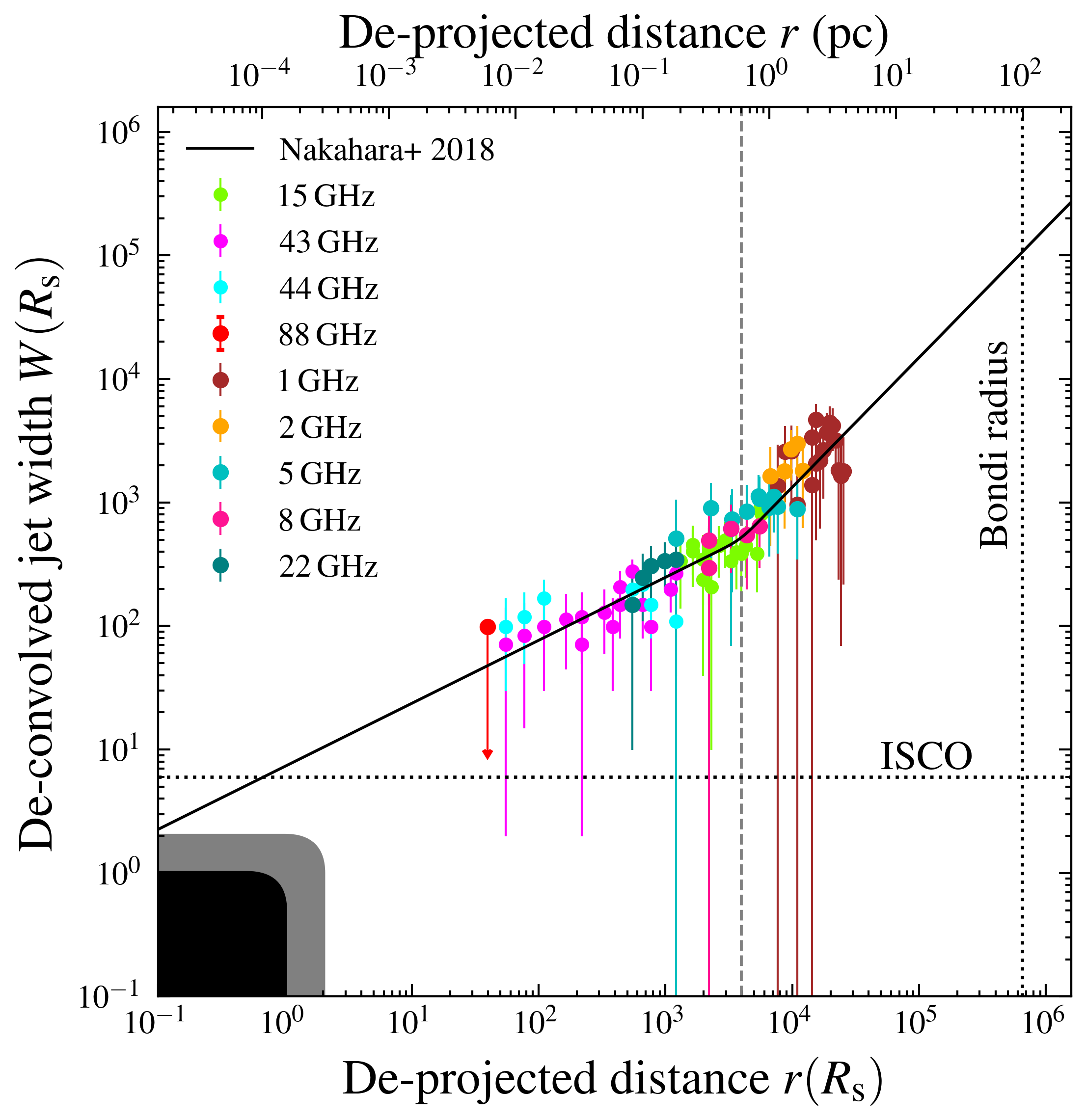}
\caption{Top: Power-law fit of the jet width versus de-projected distance (assuming a viewing angle of $63\degree$) from the core using data at 15, 43, 44 and 88\,GHz. Bottom: Same as the top panel, but including the 1, 2, 5, 8, and 22\,GHz data. The black solid line represents the radial width profile fit from \cite{nakahara_2018ApJ...854..148N}. The vertical dashed line indicates the location of the structural transition. The black and grey areas represent the size of the event horizon surface for black holes with maximum spin and no spin, respectively.}
\label{fig:inner_collimation}
\end{center}
\end{figure}

In the top panel of Figure\,\ref{fig:inner_collimation}, we present the combined results obtained from both the approaching and the receding jet. The inner width profile exhibits a simple power-law relationship, with the form $W \propto r^{0.56\pm0.07}$, where $W$ is the de-convolved jet width and $r$ denotes the de-projected distance from the black hole. This power-law relationship corresponds to a parabolic jet shape.

We also measured the width of the downstream jet based on previous multi-frequency (1.4, 2.3, 5.0, 8.4, and 22\,GHz) VLBA observations \cite{nakahara_2018ApJ...854..148N}. We re-imaged the source and determined the jet width as in Section\,\ref{sec:transverse structure}. The results are shown in the bottom panel of Figure\,\ref{fig:inner_collimation}. With these multi-frequency jet width measurements, the width profile clearly show a transition from parabolic to conical shape.
We note that this transition is in good agreement with the broken power-law function fitted by \citet{nakahara_2018ApJ...854..148N} (see their Eq.(1) and Table 2)
\footnote{We shifted the fitting line to account for the different black hole masses used in their study ($4.9\times10^{8} M_{\sun}$) and our study ($1.62\times10^{9} M_{\sun}$).}. We emphasize that the jet collimation is already completed at sub-parsec scales, with the transition location of $\sim$0.61 pc or $4\times10^{3} R_{\rm s}$ being significantly smaller than the Bondi radius \citep[99.2 pc or $r_{\rm B}\sim 6.5\times10^{5} R_{\rm s}$,][]{Balmaverde_2008A&A...486..119B}
\footnote{In their original paper, the calculated Bondi radius was 32 pc, based on a black hole mass of $5.25\times10^{8} M_{\sun}$, which is 3.1 times smaller than the mass we used. Therefore, we adopted a Bondi radius of 99.2 pc ($r_{\rm B}\propto M_{\rm BH}$.)}.

\section{Discussion} 
\label{sec:Discussions} 
In this study, we presented the first multi-epoch kinematic analysis of the NGC\,4261 jet. Previous studies by \citet{piner_2001AJ....122.2954P} reported an apparent speed of $0.83\pm0.11$ mas/year at about 5--6 mas from the core based on two-epoch observations. By combining this value with the jet/counter-jet brightness ratio and the spectral index, they derived a jet viewing angle of $63\degree\pm3\degree$. We found that the apparent jet speeds in our study are consistent with the previous results. The derived viewing angle by \citet{piner_2001AJ....122.2954P} also falls within our constrained range. 
In addition, with the caveat that the measured proper motions should not be over-interpreted, the increase in the apparent speeds from 0.31\,$c$ to 0.59\,$c$ suggests that the jet may be undergoing acceleration. We note that this acceleration is observed on the sub-parsec scale (de-projected), largely coinciding with the jet collimation region. Future high-resolution and high-cadence observations will allow a more detailed study of this jet acceleration.

Compared to previous studies \citep{nakahara_2018ApJ...854..148N}, we provide a more comprehensive examination of the innermost jet structure using the high-sensitivity data. We confirm that the innermost jet exhibits a parabolic shape. Notably, we found that the transition location of the width profile (0.61 pc or $\sim4\times10^{3} R_{\rm s}$) is significantly smaller than the corresponding Bondi radius (99.2 pc or $\sim6.5\times10^{5} R_{\rm s}$). Interestingly, this behavior is similar to that observed in the nearby radio source, NGC\,315, where the jet transition location is also at a significantly smaller distance from the core than the Bondi radius \citep{boccardi_2021A&A...647A..67B,park_2021ApJ...909...76P}.

Similar to NGC\,315, we propose that the shape transition in NGC\,4261 is influenced by external pressure from the surrounding medium. Following the discussions on NGC\,315 by \cite{boccardi_2021A&A...647A..67B}, we investigate potential sources of the external pressure in NGC\,4261. One possibility is the ADAF itself. Previous observations and theoretical models have shown that the ADAF model is crucial in explaining the X-ray emission in NGC\,4261 \citep{Gliozzi_2003A&A...408..949G,Nemmen_2014MNRAS.438.2804N}. And it is also suggested that the ADAF is truncated by an outer thin disk at a location of $\sim10^{3}-10^{4}R_{\rm s}$ \citep{Gliozzi_2003A&A...408..949G, Nemmen_2014MNRAS.438.2804N}. Notably, this truncation location is comparable to the location of the jet shape transition. Therefore, the parabolic jet profile may be initially collimated by the thick ADAF itself.

Alternatively, the external pressure may be provided by a non-relativistic disk wind rather than the ADAF \citep[e.g.,][]{Blandford_2022MNRAS.514.5141B}. The disk wind is believed to originate from the ADAF, and its role in shaping the parabolic geometry has been studied in M\,87 \citep[e.g.,][]{Globus_2016MNRAS.461.2605G,Nakamura_2018ApJ...868..146N}. In the case of NGC\,4261, considering reasonable conditions \citep{boccardi_2021A&A...647A..67B,Globus_2016MNRAS.461.2605G}, the wind may efficiently collimate and confine the jet.

On the other hand, the transition in the internal flow, from a magnetically dominated to a particle-dominated regime, could also account for the observed jet profile transition. A recent semi-analytical model proposed by \cite{Kovalev_2020MNRAS.495.3576K} supports this idea. According to their model, the jet profile transition can occur under the influence of a single power-law external pressure profile. Importantly, the location of the transition point in the profile is closely tied to the initial magnetization of the jet and can lie within the region well below the Bondi radius \citep[see Figure 8 in][]{Kovalev_2020MNRAS.495.3576K}. Based on these, we propose that the initial confinement of the jet is also possibly due to the magnetic pressure that is dominated in a region far below the Bondi radius.

Lastly, it is interesting to note that the jet width in NGC\,4261 appears to be comparable to that in M\,87 on the same physical scales. This contradicts the previous findings by \cite{nakahara_2018ApJ...854..148N}, who found that jet width in NGC\,4261 is much larger than that in M\,87. However, this can be attributed to the use of a smaller black hole mass in their study.

\section{Summary}\label{sec:summary}
In this paper, we presented multi-frequency VLBI studies of the kinematics and collimation of the two-sided jets in NGC\,4261 on sub-parsec scales. Our findings are summarized as follows:

\begin{enumerate}
\item 
We obtained VLBI images of NGC\,4261 at 15, 43, 44 and 88\,GHz. At 43 and 44\,GHz, we observed a more extended double-sided structure compared to previous studies. At 88\,GHz, we obtained a clear image of the nuclear structure at a scale as small as 100 $R_{\rm s}$. We found the core size at 88\,GHz is 0.09 mas and the brightness temperature is $\sim 1.3\times10^{9}$\,K.

\item 
We measured proper motions in both the approaching and receding jets on sub-parsec scales. The measured apparent speeds in the approaching jet range from $0.31\pm0.14\,c$ to $0.59\pm0.40\,c$. The increase in apparent speeds with distance from the core suggests an acceleration of the jet, which will need to be confirmed by future observations. Furthermore, we also observed a jet speed of $0.32\pm0.14\,c$ in the counter-jet.

\item
Using the measured apparent velocity and the jet-to-counter-jet brightness ratio, we constrained the jet viewing angle to between $54\degree$ and $84\degree$. We also found that the intrinsic speed is between $0.30\,c$ and $0.55\,c$. Combining these results with the jet collimation profile suggests that the jet acceleration region possibly coincides with the jet collimation region.

\item
We found a parabolic shape for the upstream jet on both sides, described by $W \propto r^{0.56\pm0.07}$. We emphasize that the jet collimation is already completed at sub-parsec scales. Combining our findings with previous studies, we found that the transition location of the jet structure (0.61 pc or $\sim4\times10^{3} R_{\rm s}$) is significantly smaller than the corresponding Bondi radius (99.2 pc or $\sim6.5\times10^{5} R_{\rm s}$). This behavior is similar to what has been observed in NGC\,315. Like NGC\,315, we interpret this behavior as the initial confinement of the jet by the external pressure exerted by either the geometrically thick, optically thin ADAF or the disk wind launched from it. Alternatively, the shape transition may also be explained by the internal flow transition from a magnetically dominated to a particle-dominated regime.

\end{enumerate}

\begin{acknowledgments}
We thank the anonymous referee for helpful comments and suggestions. This work was supported by the Key Program of the National Natural Science Foundation of China (grant no. 11933007), the Key Research Program of Frontier Sciences, CAS (grant no. ZDBS-LY-SLH011), the Shanghai Pilot Program for Basic Research, Chinese Academy of Sciences, Shanghai Branch (JCYJ-SHFY-2022-013) and the Max Planck Partner Group of the MPG and the CAS. The Very Long Baseline Array is operated by the National Radio Astronomy Observatory, a facility of the National Science Foundation, operated under cooperative agreement by Associated Universities, Inc.
\end{acknowledgments}

\bibliographystyle{aasjournal}
\bibliography{references}
\end{document}